\documentclass[runningheads]{llncs}
\usepackage{graphicx}
\usepackage[T1]{fontenc}
\usepackage{bbm}
\usepackage{amsmath}
\usepackage{amsfonts}
\usepackage{multirow}
\usepackage{hyperref}
\usepackage{enumitem}
\usepackage[table,xcdraw]{xcolor}
\usepackage{tabularx} % Add this in your preamble
\usepackage[table]{xcolor} % Add this in your preamble if not already added
\usepackage{cite}
\usepackage{xcolor}

\usepackage{bbold}
\usepackage{float}

\usepackage{wrapfig}

\begin{document}

\title{Biophysics Informed Pathological Regularisation \\ for Brain Tumour Segmentation}

\titlerunning{Biophysics Informed Pathological Regularisation}

% \institute{}
\author{Lipei Zhang  \and
Yanqi Cheng \and
Lihao Liu \and \\
Carola-Bibiane Schönlieb \and 
Angelica I Aviles-Rivero}

\authorrunning{L. Zhang et al.}
\institute{Department of Applied Mathematics and Theoretical Physics, \\ University of Cambridge, UK \\
\email{\{lz452,yc443,ll610,cbs31,ai323\}@cam.ac.uk}}

\maketitle

\begin{abstract}
Recent advancements in deep learning have significantly improved brain tumour segmentation techniques; however, the results still lack confidence and robustness as they solely consider image data without biophysical priors or pathological information. Integrating biophysics-informed regularisation is one effective way to change this situation, as it provides an prior regularisation for automated end-to-end learning. In this paper, we propose a novel approach that designs brain tumour growth Partial Differential Equation (PDE) models as a regularisation with deep learning, operational with any network model. Our method introduces tumour growth PDE models directly into the segmentation process, improving accuracy and robustness, especially in data-scarce scenarios. This system estimates tumour cell density using a periodic activation function. By effectively integrating this estimation with biophysical models, we achieve a better capture of tumour characteristics. This approach not only aligns the segmentation closer to actual biological behaviour but also strengthens the model's performance under limited data conditions. We demonstrate the effectiveness of our framework through extensive experiments on the BraTS 2023 dataset, showcasing significant improvements in both precision and reliability of tumour segmentation.

\keywords{Glioma \and Segmentation \and Partial Differential Equations \and Deep Learning  \and Representation Regularisation.}
\end{abstract}

\section{Introduction}
Glioblastoma, the most aggressive brain cancer, represents 14.3\% of primary malignant central nervous system tumours. It exhibits rapid, heterogeneous growth that complicates detection and treatment.  Thus often leading to poor prognosis despite advanced interventions~\cite{low2022primary,who2016}. Magnetic Resonance Imaging (MRI) plays a crucial role in diagnosing brain tumours. It provides high-resolution images that are essential for accurate tumour margin segmentation. These images also guide treatment decisions across various modalities, such as T1, T1-weighted with contrast (T1w-gd), T2-weighted (T2w), and T2 Fluid Attenuated Inversion Recovery (FLAIR)~\cite{bakas2017advancing}. Advances in deep learning, particularly with introduction of the U-Net architecture~\cite{ronneberger2015u}, have significantly improved segmentation in different imaging modalities~\cite{cciccek20163d,milletari2016v,oktay2018attention,alom2018recurrent,liu2020psi,isensee2021nnu}. The integration of U-Net and Transformer models~\cite{hatamizadeh2021swin,liu2022pc,hatamizadeh2022unetr}, utilising long-range attention mechanisms, show improved accuracy in the Brain Tumour Segmentation (BraTS) Challenge~\cite{baid2021rsna}.

However, the limited data availability in biology and medicine highlights the challenges of the generalisability issues arising from the limited availability of medical data. This underscores the importance of incorporating domain-specific knowledge into deep learning models to enhance their utility in medical imaging. Physics-Informed Neural Networks (PINNs) offer a promising approach to overcome generalisability issues. They integrate boundary conditions and partial differential equations (PDEs) into the learning process. It has enabled their application across various fields.  For example, it is used in elasticity reconstruction~\cite{ragoza2023physics}, Alzheimer’s disease analysis~\cite{song2020physics}, and glioma progression estimation~\cite{meaney2023deep}. However, the effort in glioma has primarily focused on parameter estimation. Previous efforts have not succeeded in directly integrating physics-informed learning with automated segmentation models. The integration could leverage the inherent structure and dynamics of biological systems. This approach has the potential to yield more accurate and biologically plausible segmentation results. This method could connect data-driven models with model-driven models that include specific expert knowledge in their learning.

In this work, we propose a novel method that designs biophysics-informed regularisation to improve brain tumor segmentation through deep learning. This approach embeds pathological insights into tumour growth dynamics, enhancing segmentation accuracy and robustness. Our contributions are:

\begin{enumerate}[itemsep=1pt,topsep=4pt,parsep=5pt]
    \item A unified framework that integrate deep learning with brain tumour growth PDE models through a biophysics-informed regularisation, enhancing segmentation precision and robustness across various scenarios.
    \item Introducing MRI-driven tumour cell density estimator estimates along with biophysics-informed regularisation, achieving segmentations that reflect real tumour growth mechanism. In this estimator, a designed periodic activation function in the high-level feature extracting layers captures the non-linear dynamics of the biophysical model, enabling detailed signal and spatial derivative learning for calculation of biophysics-informed regularisation term.
    \item These innovations have yielded exceptional segmentation results on the BraTS 2023 dataset, proving effective across various architectures, dataset sizes, missing modalities and combinations with other losses.
\end{enumerate}

\section{Proposed Method}
\paragraph{\textbf{Brain tumour segmentation.}} In this problem, we aim to segment 4 regions: normal region ($y_0$), tumour core (TC, $y_1$), and whole tumour (WT, $y_2$), enhancing tumour (ET, $y_3$), in a given MRI data with T1 ($i_0$), T1Gd ($i_1$), T2 ($i_2$), and FLAIR ($i_3$) volumes. The segmentation model can be defined as \( f_{\theta} 
% : I \rightarrow \Sigma 
\) such that the probability of each class of \( P \) can be obtained as \( P = f_{\theta}(I) \), where $I = \{i_0, i_1, i_2, i_3\}$, $P = \{p_0, p_1, p_2, p_3\}$, and $Y = \{y_0, y_1, y_2, y_3\}$ indicates one-hot mask of each subregion.
% \vspace{-10pt}
\begin{figure*}[t!]
% \begin{wrapfigure}{r}{0.62\textwidth}
% \vspace{-20pt}
 \setlength{\abovecaptionskip}{-0.cm}
    \centering
    \includegraphics[scale=0.43]{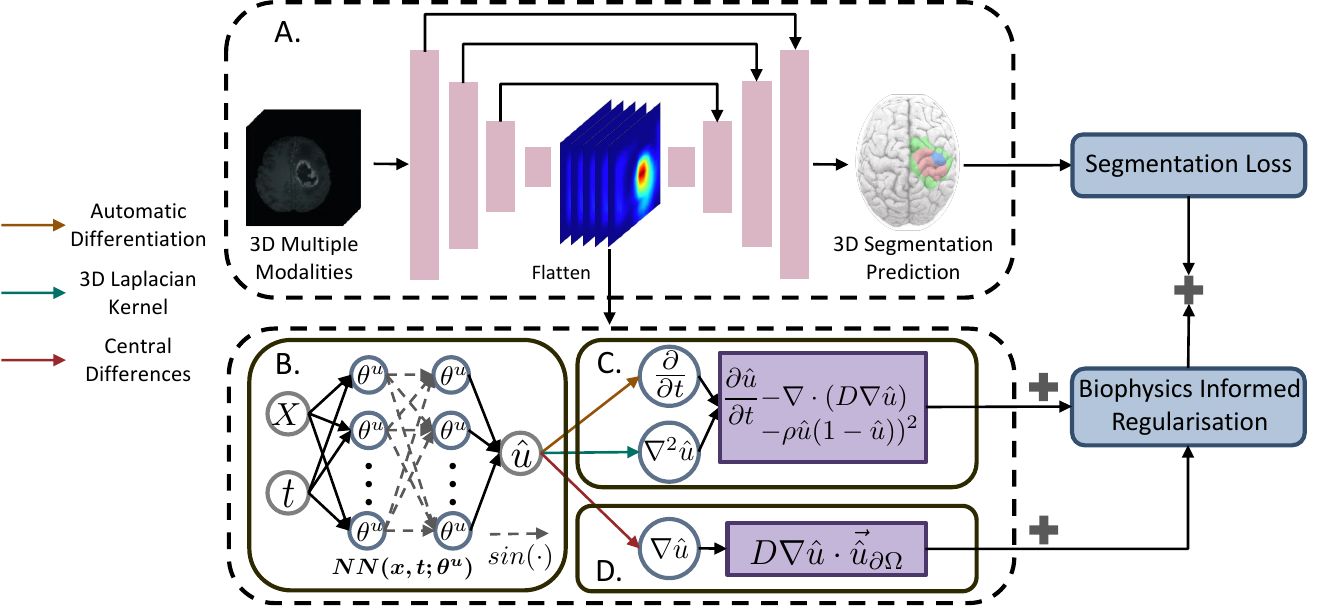}
    \caption{The Biophysics-informed optimisation for segmentation. A. Main structure for brain tumour segmentation. B. Tumour cell density estimator (the flattened feature map will be concatenated with assumed time matrix T). C. Calculation of PDE loss. D. Calculation of boundary loss.} 
\end{figure*}
% \end{wrapfigure}
% \vspace{-10pt}
The function \( f_{\theta} \) is usually nonlinear and \( \theta \) is a large vector of parameters. The learning phase selects \( \theta \) in order to minimise a loss function \( \mathcal{L} \) that measures the accuracy of the predicted segmentation \( f_{\theta}(I) \). This overall training flow is shown in Fig. 1(a). The parameters $\theta$ are obtained by minimising a loss function plus a regularisation term. Consider a dice loss with supervised manner, we optimise the network by:
% \vspace{-8pt}
\begin{equation}
\hspace{-0.2cm}
\arg \min_{\theta} \mathcal{L}(I, Y, \theta) = \arg \min_{\theta} \Bigg( \underbrace{\sum_{n \in \{0,1,2,3\}} \left(1 - \frac{2 y_n p_n + 1}{y_n + p_n + 1}\right)}_\text{Segmentation Loss} + \underbrace{\vphantom{\sum_{n \in \{0,1,2,3\}}} R_{\theta}(f_{\theta}(I))}_\text{Regularisation}\Bigg)
% \theta^* = 
% \hspace{-0.05cm}
% \arg \min_{\theta} \mathcal{L}(I, Y, \theta) = \arg \min_{\theta} \left\{ \sum_{n \in \{0,1,2,3\}} \left(1 - \frac{2 y_n p_n + 1}{y_n + p_n + 1}\right) + R_{\theta}(f_{\theta}(I)) \right\}\\
% 
% \hspace{0.7cm} \underbrace{\qquad \qquad \qquad \qquad \qquad \quad \quad}_\text{Segmentation Loss} \,  \underbrace{\qquad \qquad }_\text{Regularisation Term}
% \vspace{-0.2cm}
\end{equation}
\vspace{-25pt}
% \vspace{-0.42cm}
\\
\paragraph{\textbf{Biophysics-informed optimisation.}} Exploring effective regularisation terms $R_\theta (\cdot)$ is essential to improve model robustness. Unlike current regularisation approaches based on VAE \cite{myronenko20193d}, we propose that incorporating biophysical models of tumour growth model into the training process as an more explainable learning bias. It enables the segmentation model to capture more accurate pathological features representative of tumour areas. To this end, we introduce a biophysics-informed regularisation module designed as a plug-and-play component. This module integrates two key elements: additional fully connected layers with periodic activation functions for pathological information extraction, and a combination of proliferation and diffusion PDEs alongside boundary conditions. This integration not only improves the segmentation's ability to capture tumour details by embedding biophysical regularisation directly into the end-to-end optimisation process but also ensures the model's outputs align more closely with actual biological behaviours. The overall objective loss, balanced by weights $\lambda_1$ and $\lambda_2$, is formulated as:
% \vspace{-5pt}
\begin{equation}
\mathcal{L}_{\text{Total}} = \mathcal{L}_{\text{Dice}} +~\underbrace{\lambda_{1}\mathcal{L}_{\text{PDE}} + \lambda_{2}\mathcal{L}_{\text{BC}}}_\text{\hspace{-1.1cm}Biophysics Informed Regularisation \hspace{-1.1cm}}
\end{equation}
% \vspace{-10pt}

\paragraph{\textbf{Tumour cell density estimator.}} In this first part of this module, we introduce a neural network component, \( NN(x, t; \theta^{u}) \), which employs multiple fully connected layers with periodic activation functions to disentangle high-level feature maps from middle stage to potential tumour cell density $\hat{u}$. At these stages, the feature maps primarily highlight pathological features, focusing on tumour regions. This architecture offers a significant advantage, that it learns the non-linear aspects of governing PDEs indirectly, thus endowing the model with the capability to capture detailed signals and spatial derivatives without specifying them explicitly. 

The process begins with a set of feature maps $(B, C, H, W, D)$, where $B$ represents batch size, $C$ is channel number and $H, W, D$ denote height, width and depth, respectively. These maps are flattened across $H, W, D$ dimensions to form a matrix $X$ with dimensions  $(B, C, H\times W\times D)$. This procedure ensures efficient sequential processing and maintains the independence of the information of each channel. This approach, different from SIREN's coordinate consideration \cite{siren}, incorporates an assumed temporal dimension $T$, viewing each segmentation optimisation step as a discrete moment in time. The temporal dimension $T$, a matrix of the same shape as $X$, but filled with the assumed time step $t$, is concatenated with $X$. Time embedding is crucial for simulating tumour dynamics during training, ensuring the model's output aligns with expected growth dynamics. It also facilitates the calculation of the first-order time derivative, integrating temporal dynamics into our model. The concatenated features $y = (X, T)=((x_{1},t_{1}) (x_{2},t_{2}), ..., (x_{t},t_{C}))$ with size of $(B, C, H\times W\times D \times 2)$, with $x_c$ and $t_c$ represent dimensional and temporal matrix on the $c^{th}$ channel respectively, undergo transformation as follows:
% \vspace{-5pt}
\begin{equation} \label{sine}
    \hat{u} = \Gamma(y) = W_n(\gamma_{n-1} \circ \gamma_{n-2} \circ \cdots \circ \gamma_0)(y) + b_n, \quad y_i \mapsto \gamma_i(y_i) = \sin(W_iy_i + b_i)
\end{equation}

\noindent
where, \(\gamma_i : \mathbb{R}^{M_i} \rightarrow \mathbb{R}^{N_i}\) denotes the \(i^{th}\) layer of the transformation. It comprises the affine transforms defined by the weight matrix \(W_i \in \mathbb{R}^{N_i \times M_i}\), and the biases \(b_i \in \mathbb{R}^{N_i}\) utilised on the input \(y_i \in \mathbb{R}^{M_i}\), followed by the sine nonlinearity. This process extracts the tumour cell density $\hat{u}$, facilitating the calculation of tumour proliferation and diffusion PDEs. During the inference phase, this module can be pruned to avoid any increase in processing time.

\paragraph{\textbf{Calculation of biophysics-informed regularisation.}} Tumour cell density proliferation and diffusion across the brain is predominantly based on the reaction-diffusion equation~\cite{le2015bayesian}, the model incorporates logistic growth to represent cell proliferation and enforces Neumann boundary conditions to model tumour boundaries within the brain domain $\Omega$,
% \vspace{-5pt}
\begin{equation} \label{pde}
    \begin{gathered}
    \frac{\partial u}{\partial t} = \underbrace{\nabla \cdot (d \nabla u)}_{\text{Diffusion}} + \underbrace{f(u)}_{\text{Reaction}} \\
    f(u) = \rho u\left(1-u\right), \quad
    d \nabla u \cdot \vec{u}_{\partial \Omega}=0
    \end{gathered}
\end{equation}
% \vspace{-11pt}

These equations describes the evolution of the tumour cell density \( u \), which infiltrates neighbouring tissues with a diffusion tensor \( d \), and proliferates according to the law defined with \( f(u) \). We consider the reaction term with a logistic growth term with a net proliferation rate \( \rho \). Here, \( d \) and \( \rho \) have same dimension with \( u \).
The $\frac{\partial u}{\partial t}$ with respect to time $t$ is computed through automatic differentiation due to $t$ embedded with input $x$. The predicted cell density vectors \( u \) are reshaped back to 3D format $(B, C, H, W, D)$. $F_{\Delta}(u)=\nabla \cdot (d \nabla u)$, is approximated using a 3D Laplacian kernel ($K$). This kernel is more efficient than traditional central finite difference method for preserving spatial information when computing the second derivative and boundary condition related to location. 
% \vspace{-5pt}
{\small
\begin{equation} \label{kernal}
K =
% \[
\left[
\begin{array}{ccc}
\left.
\begin{matrix}
0 & 0 & 0 \\
0 & 1 & 0 \\
0 & 0 & 0
\end{matrix}
\right\} & 
\left.
\begin{matrix}
0 & 1 & 0 \\
1 & -6 & 1 \\
0 & 1 & 0
\end{matrix}
\right\} & 
\left.
\begin{matrix}
0 & 0 & 0 \\
0 & 1 & 0 \\
0 & 0 & 0
\end{matrix}
\right\}
\end{array}
\right]
% \]
\end{equation}
}
% \vspace{-10pt}

After considering all voxels with total number of $N_{biophy} = H\times W \times D$ in feature maps, the PDE loss term takes the following form:
% \vspace{-8pt}
\begin{equation} \label{pde_loss}
\mathcal{L}_{\text{PDE}} =\Big\|\frac{1}{N_{biophy}} \sum_{i=1}^{N_{biophy}}\left(\frac{\partial u_{i}}{\partial t_i}-\nabla \cdot (D \nabla u_{i}) - \rho u_{i}(1-u_{i})\right)^2 \Big\|
\end{equation}
% \vspace{-10pt}

$||\cdot||$ represents the norm across channels and batches. Additionally, boundary constraints are specified to model the diffusion process accurately, ensuring the diffusion remains non-negative and aligns with the spatial domain of the brain within the feature maps. Here, this can be describe by the following formula:
{\small
\begin{equation} \label{bc}
\begin{aligned}
\quad 0 \leq x \leq H, \quad &0 \leq y \leq W,  \quad0 \leq  z \leq D\\
u_{x}(0, y, z) &= 0, \quad u_{x}(H, y, z) = 0 \\
u_{y}(x, 0, z) &= 0, \quad u_{y}(x, W, z) = 0 \\
u_{z}(x, y, 0) &= 0, \quad u_{z}(x ,y, D) = 0
\end{aligned}
\end{equation}
}
% \vspace{-11pt}

The corresponding boundary constraints take the form, with $\mathbb{1}$ indicates the indicator function:
% \vspace{-7pt}
\begin{equation} \label{bc_loss}
% \begin{aligned}
% \Scale[0.9]{
\mathcal{L}_{\text{BC}} = \Big\| d %\cdot
\sum_{\substack{x=0}}^{H}\sum_{\substack{y=0}}^{W}\sum_{\substack{z=0}}^{D} \Big[
\frac{ {\mathbb{1}}_{\{0,H\}}(x) }{WD} u_{x}^2(y,z)
+ \frac{{\mathbb{1}}_{\{0,W\}}(y)}{HD} 
  u_{y}^2(x,z)
+\frac{{\mathbb{1}}_{\{0,D\}}(z) }{HW}
u_{z}^2(x,y) \Big] \Big\|
% }
% \end{aligned}
\end{equation}

\section{Experiment}
\paragraph{\textbf{Datasets.}} Our evaluations were applied on the BraTS 2023 dataset~\cite{baid2021rsna}, consisting of 1251 cases within the official training partition. Each case comprises mpMRI scans with four modalities: T1, T1c, T2, and FLAIR, each having dimensions of $240\times240\times155$. Three tumor subregions---enhancing tumour (ET), tumour core (TC), and whole tumour (WT)---were labelled by neuroradiologists. Given the absence of labels in the official validation and testing sets, the 1251 cases was split into training, validation, and test sets with a ratio of 7:1:2. During training, the 4-channel MRI volumes centred on the tumour were cropped to $128\times128\times128$ voxel patches. To mitigate overfitting, augmentation techniques were employed. Z-score standardisation was performed on non-zero voxels across each channel, with outlier values being clipped. For inference, same centred cropping on the tumour of 4-channel MRI volumes was conducted, complemented by test-time augmentation (TTA) techniques~\cite{shanmugam2021better} and overlapped sliding window inference~\cite{isensee2021nnu}. Other preprocessing setups aligned with~\cite{carre2021automatic}.

\paragraph{\textbf{Main experimental Details and compared methods.}}
All experiments, developed on MONAI v1.3.0~\cite{maproject} and Pytorch 2.1.0, was executed on an Nvidia A10 GPU with 24GB memory with automatic mixed precision training. The number of epoch was 175 epochs and batch size was set to 1. We used the Ranger 2020 optimiser \cite{zhang2019lookahead, liu2019variance} with a starting learning rate of $3 \times 10^{-4}$ and a cosine decay schedule. Dice Loss \cite{milletari2016v} was computed both batch-wise and channel-wise, unweighted, alongside a consistent loss weight for PDE and BC ($\lambda_1$ and $\lambda_2$ set to 1). Final models were used for testing. Baseline comparisons included 3D UNet~\cite{ronneberger2015u}, R2-UNet~\cite{alom2018recurrent}, nn-UNet~\cite{isensee2021nnu}, UNETR~\cite{hatamizadeh2022unetr}, SegResNet, and SegResNetVAE~\cite{badrinarayanan2017segnet}, all employing single dice loss. We compared these with our proposed biophysics-informed loss, excluding SegResNetVAE for regularization comparison. More configurations are shown in appendix. The diffusion coefficient $d$ and proliferation rate $\rho$ are in the range of $[0.02,1.5]$ $mm^2/day$ and $\rho$ between $[0.002,0.2]/day$
% ^{-1}$ 
for tumour tissues respectively~\cite{le2015bayesian}. High-level feature maps, targeting tumour region, assume isotropic diffusion across voxels due to gliomas' tendency to disrupt neural pathways, leading to uniform diffusion. The input size of feature map for the tumour cell estimator was $(B \times C \times 16 \times 16 \times 16)$ to ensure computational efficiency. The $d$ and $\rho$ values randomly sampled from the specified ranges for each voxel. This random sampling introduces regularisation, ensuring biophysical priors fall within a valid distribution.

\paragraph{\textbf{Evaluation Metrics.}}
The results were compared with 2 quantitative criteria, Dice similarity coefficient (Dice)~\cite{dice1945measures} and Hausdorff Distance (HD)~\cite{huttenlocher1993comparing}, on each of the three sub-regions. We take the norm as Euclidean norm for HD. Outliers were removed using the 95th percentile (HD95) approach~\cite{carre2021automatic}.

\paragraph{\textbf{Results and Discussion.}} The segmentation results for the BraTS2023 dataset, shown in Table.~\ref{tab:main}, indicate that integrating biophysics-informed regularisation with UNet architectures leads to better accuracy than standard methods. Our approach not only improved the Dice coefficients for all tumour regions but also decreased Hausdorff distances, proving its effectiveness in enhancing segmentation precision and better geometric accuracy in tumour margin prediction. This method also showed consistency across different models, suggesting its robustness. By integrating biophysics into training, we can achieve more reliable segmentations, potentially improving clinical outcomes.

\begin{table}[htbp]
\centering 
\caption{Comparison table of UNet family with and without the
biophysics-informed regularisation on mean and standard deviation of Dice score and Hausdorff distance. 
\vspace{-5pt}
}
\label{tab:main}
\footnotesize % This command will further reduce the font size
\setlength\tabcolsep{12pt} % Adjust the space between columns if necessary
\renewcommand{\arraystretch}{1} % Reduce the space between rows
\setlength{\abovedisplayskip}{0pt}
\setlength{\belowdisplayskip}{0pt}
\begin{tabular}{l|ll}
\hline
Method             & Dice $\uparrow$              & Hausdorff Distance (mm)$\downarrow$ \\
                   & TC/WT/ET          & TC/WT/ET                \\ \hline
UNet               & 90.68/92.29/87.28 & 6.16/7.85/10.99         \\
\rowcolor[HTML]{EFEFEF} 
UNet (Biophy)      & 91.83/92.34/88.26 & 3.99/6.94/7.47          \\ \hline
R2-UNet            & 90.75/91.86/87.29 & 5.82/7.34/11.24         \\
\rowcolor[HTML]{EFEFEF} 
R2-UNet (Biophy)   & 91.02/91.91/87.53 & 4.83/7.14/9.25          \\ \hline
nn-UNet            & 91.24/92.48/87.38 & 4.44/6.95/10.85         \\
\rowcolor[HTML]{EFEFEF} 
nn-UNet (Biophy)   & 91.70/92.69/87.55 & 4.00/6.51/10.39         \\ \hline
UNet-TR            & 88.20/90.73/85.41 & 6.75/11.36/12.62        \\
\rowcolor[HTML]{EFEFEF} 
UNet-TR (Biophy)   & 89.06/91.10/85.96 & 5.97/8.98/11.12         \\ \hline
SegResNet          & 91.96/91.99/87.28 & 3.91/7.53/10.80         \\
SegResNet-VAE      & 92.00/91.02/87.02 & 3.88/6.80/10.73         \\
\rowcolor[HTML]{EFEFEF} 
SegResNet (Biophy) & 92.13/92.19/87.46 & 3.62/6.68/10.56         \\ \hline
\end{tabular}
\vspace{-5pt}
\end{table}
% \vspace{-20pt}
\begin{figure}[ht] 
\hspace{-0.6cm}
\centering
\renewcommand*{\arraystretch}{0}
\begin{tabular}{*{4}{@{}c}@{}}
\includegraphics[scale=0.25]{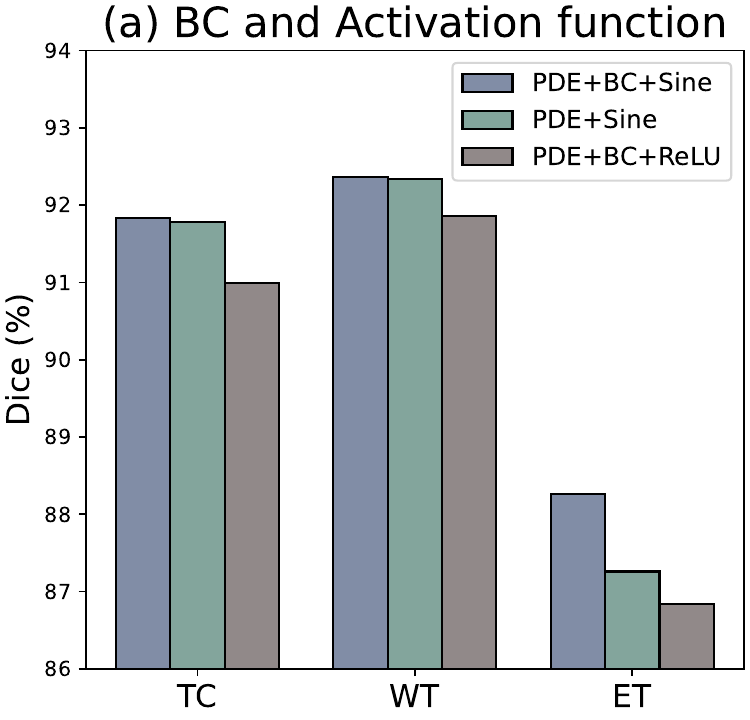}    & 
\includegraphics[scale=0.25]{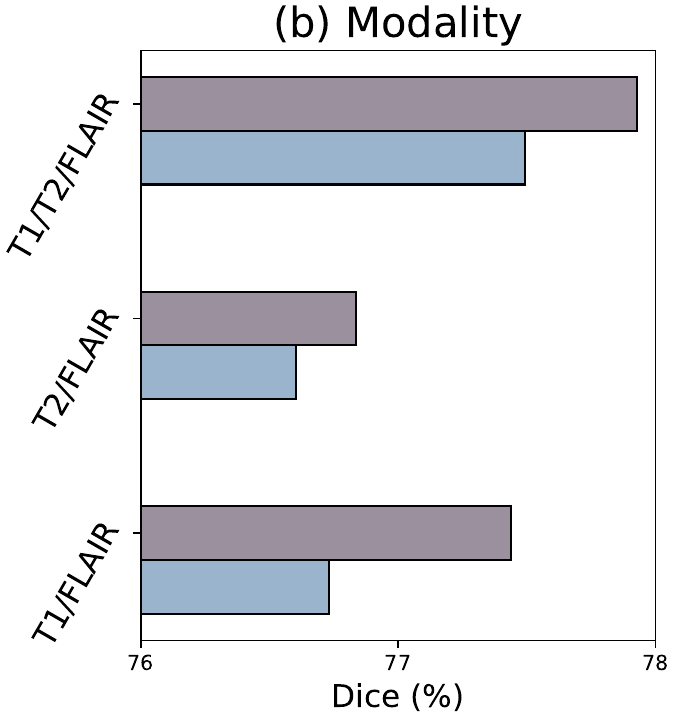}   & 
\includegraphics[scale=0.25]{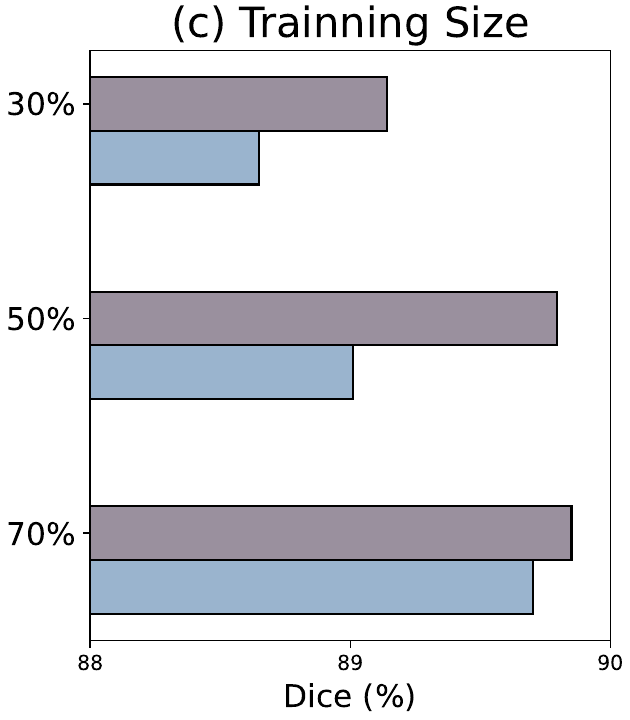}   & 
\includegraphics[scale=0.25]{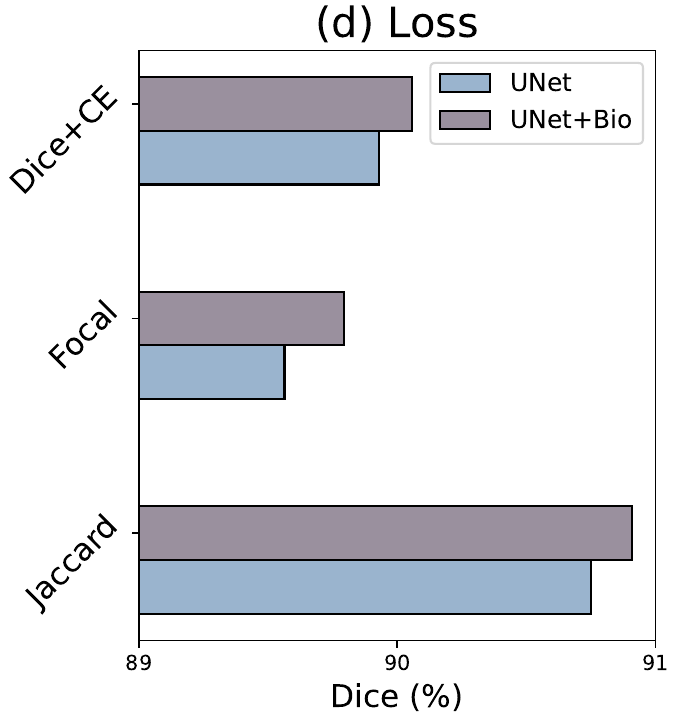}  
\end{tabular}
\caption{(a) shows the impact of activation functions and boundary conditions on the biophysics-informed UNet performance on Dice score.
(b)-(d) shows comparisons of UNet, with and without the biophysics-informed module with mean on Dice score over the 3 regions.
(b) explores performance variations using two or three modalities, (c) presents comparisons on different training set sizes, (d) compares different loss combinations.
}
\label{fig:ablation}
\end{figure}

The additional studies depicted in Fig~\ref{fig:ablation} present an comprehensive experiments of the biophysics-informed UNet model. In Fig~\ref{fig:ablation} (a), the employment of Sine over ReLU activation functions is scrutinised, revealing superior performance in identifying multiple regions, and the inclusion of boundary conditions mitigates excessive predictions in the potential infiltrated area (ET). These findings demonstrate that the biophysics-informed regularisation effectively offers a prior understanding of growth and marginal constraints. Fig~\ref{fig:ablation} (b) illustrates the model's robustness even with reduced imaging modalities, where using only T1 and FLAIR yields comparable results to including an additional T2 modality. Training size variations in Fig~\ref{fig:ablation} (c) show our efficiency in data-scarce scenarios. Integrating our module into UNet models even outperform standard UNets on larger datasets. Lastly, Fig~\ref{fig:ablation} (d) explores different loss functions, including Dice+Cross Entropy, Focal, and Jaccard Loss. The consistent enhancement of UNet's performance, due to biophysics informed regularisation, showcases the module's flexibility. This adaptability enhances network optimisation and performance of segmentation.

\begin{figure*}[ht]
    \centering
    \includegraphics[scale=0.35]
    % [width=1\textwidth]
    {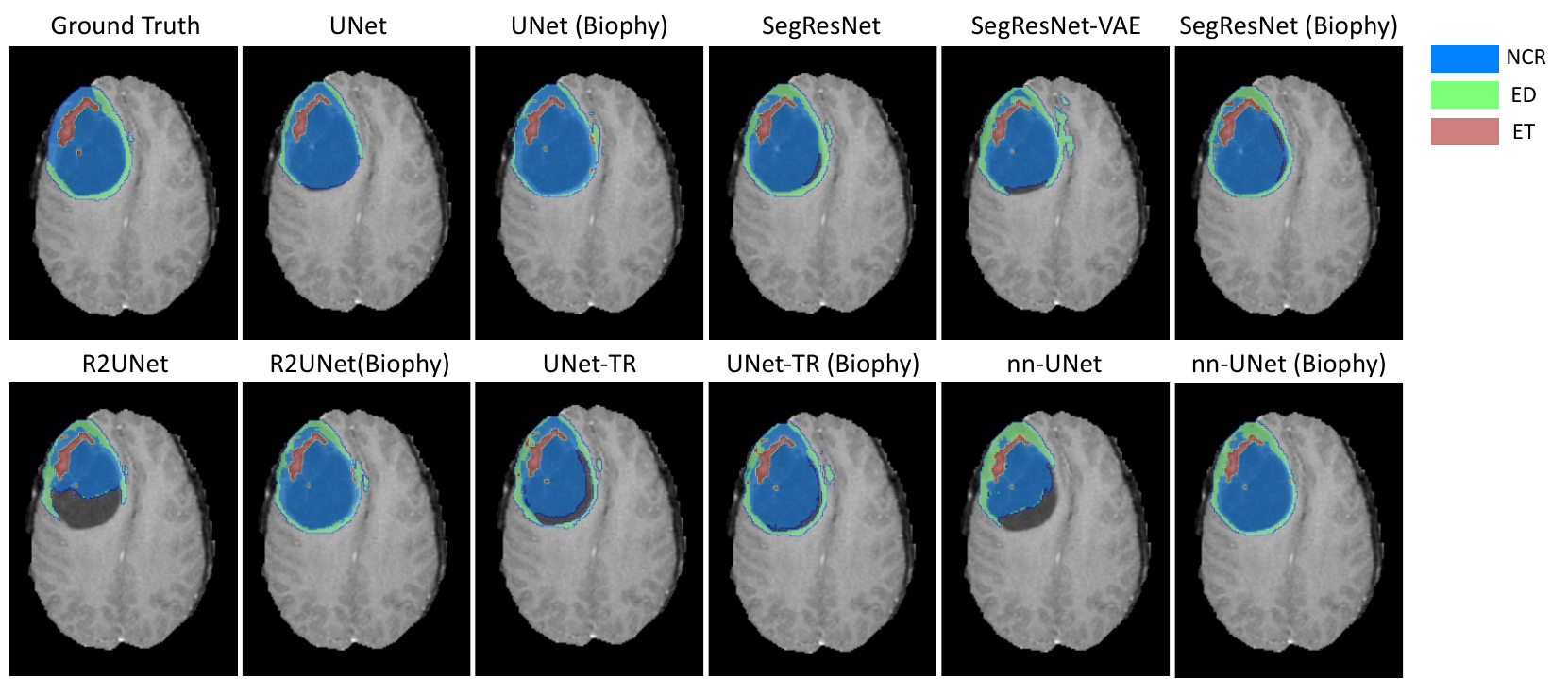} \hspace{-1cm}
    \caption{MRI tumour segmentation comparison: All enhanced by biophysics-informed regularisation. The GD-enhancing tumour (ET), the peritumoral edematous tissue (ED), and the necrotic tumour core (NCR). The WT combines red, blue, and green; the TC merges green and blue; and the ET is represented by red.} 
    \label{fig:main_comparison}
\end{figure*}
% \vspace{-25pt}
\begin{figure*}[ht]
    \centering
    \includegraphics[scale=0.5]{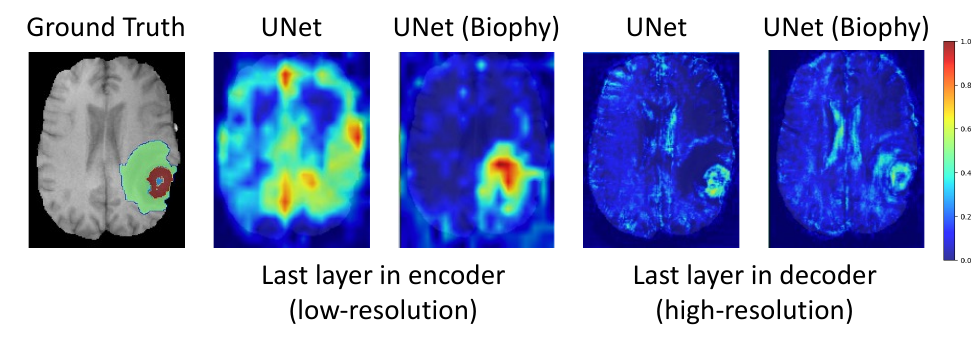} 
    \vspace{-10pt}
    \caption{Localisation in Brain Tumour Segmentation: The images compare the UNet with its biophysics-informed version using GradCAM, revealing more precise tumour localisation. The ground truth scan is shown first, followed by GradCAM visualisations of last encoder and decoder layers, with and without biophysics-informed regularisation.} 
    \label{fig:interpretability}
\end{figure*}

{Main Comparison Visualisation} in Fig.~\ref{fig:main_comparison} demonstrates that biophysics informed regularisation notably enhances tumour segmentation in various networks, yielding results that align more closely with ground truth. This method surpasses traditional dice loss and VAE regularisation by capturing tumour boundaries and structures more effectively, showcasing its potential to improve segmentation across different neural architectures.

{Interpretability Visualisation} in Fig.~\ref{fig:interpretability}, utilising Grad-CAM~\cite{selvaraju2017grad}, shows how biophysics-informed regularisation refines UNet performance. The focused activation on tumour areas, guided by biophysical tumour characteristics, leads to precise localisation and enhanced interpretability, ensuring model outputs mirror real tumour pathology for accurate medical segmentation.

\section{Conclusion}
Our study showcases the effectiveness of biophysics-informed regularisation in improving brain tumour segmentation accuracy and robustness. Integrating tumour growth dynamics with deep learning, we enhance traditional models significantly. Tested on the BraTS 2023 dataset, our method boosts various networks performance, proves robust against data scarcity and varying training losses, and sets a foundation for future exploration of domain-specific knowledge in deep learning for medical imaging, particularly in weakly supervised or unsupervised settings.

\begin{credits}
\subsubsection{\ackname} AAR gratefully acknowledges funding from the Cambridge Centre for Data-Driven Discovery and Accelerate Programme for Scientific Discovery, made possible by a donation from Schmidt Futures, ESPRC Digital Core Capability Award, and CMIH and CCIMI, University of Cambridge. This project also was supported with funding from the Oracle for Research Project Award. CBS acknowledges support from the Philip Leverhulme Prize, the Royal Society Wolfson Fellowship, the EPSRC advanced career fellowship EP/V029428/1, EPSRC grants EP/S026045/1 and EP/T003553/1, EP/N014588/1, EP/T017961/1, the Wellcome Innovator Awards 215733/Z/19/Z and 221633/Z/20/Z, CCMI and the Alan Turing Institute.

\subsubsection{\discintname}
The authors have no competing interests to declare that are relevant to the content of this article. 
\end{credits}
% \input{section/acknowledgement}

% \input{section/disclusureofinterest}

\bibliographystyle{splncs04}
\bibliography{Paper-2262}

\begin{thebibliography}{10}
\providecommand{\url}[1]{\texttt{#1}}
\providecommand{\urlprefix}{URL }
\providecommand{\doi}[1]{https://doi.org/#1}

\bibitem{alom2018recurrent}
Alom, M.Z., Hasan, M., Yakopcic, C., Taha, T.M., Asari, V.K.: Recurrent residual convolutional neural network based on u-net (r2u-net) for medical image segmentation. arXiv preprint arXiv:1802.06955  (2018)

\bibitem{badrinarayanan2017segnet}
Badrinarayanan, V., Kendall, A., Cipolla, R.: Segnet: A deep convolutional encoder-decoder architecture for image segmentation. IEEE transactions on pattern analysis and machine intelligence  \textbf{39}(12),  2481--2495 (2017)

\bibitem{baid2021rsna}
Baid, U., Ghodasara, S., Mohan, S., Bilello, M., Calabrese, E., Colak, E., Farahani, K., Kalpathy-Cramer, J., Kitamura, F.C., Pati, S., et~al.: The rsna-asnr-miccai brats 2021 benchmark on brain tumor segmentation and radiogenomic classification. arXiv preprint arXiv:2107.02314  (2021)

\bibitem{bakas2017advancing}
Bakas, S., Akbari, H., Sotiras, A., Bilello, M., Rozycki, M., Kirby, J.S., Freymann, J.B., Farahani, K., Davatzikos, C.: Advancing the cancer genome atlas glioma mri collections with expert segmentation labels and radiomic features. Scientific data  \textbf{4}(1),  1--13 (2017)

\bibitem{carre2021automatic}
Carr{\'e}, A., Deutsch, E., Robert, C.: Automatic brain tumor segmentation with a bridge-unet deeply supervised enhanced with downsampling pooling combination, atrous spatial pyramid pooling, squeeze-and-excitation and evonorm. In: International MICCAI Brainlesion Workshop. pp. 253--266. Springer (2021)

\bibitem{cciccek20163d}
{\c{C}}i{\c{c}}ek, {\"O}., Abdulkadir, A., Lienkamp, S.S., Brox, T., Ronneberger, O.: 3d u-net: learning dense volumetric segmentation from sparse annotation. In: Medical Image Computing and Computer-Assisted Intervention--MICCAI 2016: 19th International Conference, Athens, Greece, October 17-21, 2016, Proceedings, Part II 19. pp. 424--432. Springer (2016)

\bibitem{dice1945measures}
Dice, L.R.: Measures of the amount of ecologic association between species. Ecology  \textbf{26}(3),  297--302 (1945)

\bibitem{hatamizadeh2021swin}
Hatamizadeh, A., Nath, V., Tang, Y., Yang, D., Roth, H.R., Xu, D.: Swin unetr: Swin transformers for semantic segmentation of brain tumors in mri images. In: International MICCAI Brainlesion Workshop. pp. 272--284. Springer (2021)

\bibitem{hatamizadeh2022unetr}
Hatamizadeh, A., Tang, Y., Nath, V., Yang, D., Myronenko, A., Landman, B., Roth, H.R., Xu, D.: Unetr: Transformers for 3d medical image segmentation. In: Proceedings of the IEEE/CVF winter conference on applications of computer vision. pp. 574--584 (2022)

\bibitem{huttenlocher1993comparing}
Huttenlocher, D.P., Klanderman, G.A., Rucklidge, W.J.: Comparing images using the hausdorff distance. IEEE Transactions on pattern analysis and machine intelligence  \textbf{15}(9),  850--863 (1993)

\bibitem{isensee2021nnu}
Isensee, F., Jaeger, P.F., Kohl, S.A., Petersen, J., Maier-Hein, K.H.: nnu-net: a self-configuring method for deep learning-based biomedical image segmentation. Nature methods  \textbf{18}(2),  203--211 (2021)

\bibitem{le2015bayesian}
L{\^e}, M., Delingette, H., Kalpathy-Cramer, J., Gerstner, E.R., Batchelor, T., Unkelbach, J., Ayache, N.: Bayesian personalization of brain tumor growth model. In: Medical Image Computing and Computer-Assisted Intervention--MICCAI 2015: 18th International Conference, Munich, Germany, October 5-9, 2015, Proceedings, Part II 18. pp. 424--432. Springer (2015)

\bibitem{liu2020psi}
Liu, L., Hu, X., Zhu, L., Fu, C.W., Qin, J., Heng, P.A.: $\psi$-net: Stacking densely convolutional lstms for sub-cortical brain structure segmentation. IEEE transactions on medical imaging  \textbf{39}(9),  2806--2817 (2020)

\bibitem{liu2022pc}
Liu, L., Huang, Z., Li{\`o}, P., Sch{\"o}nlieb, C.B., Aviles-Rivero, A.I.: Pc-swinmorph: Patch representation for unsupervised medical image registration and segmentation. arXiv preprint arXiv:2203.05684  (2022)

\bibitem{liu2019variance}
Liu, L., Jiang, H., He, P., Chen, W., Liu, X., Gao, J., Han, J.: On the variance of the adaptive learning rate and beyond. arXiv preprint arXiv:1908.03265  (2019)

\bibitem{who2016}
Louis, D.N., Perry, A., Reifenberger, G., Von~Deimling, A., Figarella-Branger, D., Cavenee, W.K., Ohgaki, H., Wiestler, O.D., Kleihues, P., Ellison, D.W.: The 2016 world health organization classification of tumors of the central nervous system: a summary. Acta neuropathologica  \textbf{131}(6),  803--820 (2016)

\bibitem{low2022primary}
Low, J.T., Ostrom, Q.T., Cioffi, G., Neff, C., Waite, K.A., Kruchko, C., Barnholtz-Sloan, J.S.: Primary brain and other central nervous system tumors in the united states (2014-2018): A summary of the cbtrus statistical report for clinicians. Neuro-oncology practice  \textbf{9}(3),  165--182 (2022)

\bibitem{maproject}
Ma, N., et~al.: Project-monai/monai: 1.3. 0. zenodo (2023)

\bibitem{meaney2023deep}
Meaney, C., Das, S., Colak, E., Kohandel, M.: Deep learning characterization of brain tumours with diffusion weighted imaging. Journal of Theoretical Biology  \textbf{557},  111342 (2023)

\bibitem{milletari2016v}
Milletari, F., Navab, N., Ahmadi, S.A.: V-net: Fully convolutional neural networks for volumetric medical image segmentation. In: 2016 fourth international conference on 3D vision (3DV). pp. 565--571. Ieee (2016)

\bibitem{myronenko20193d}
Myronenko, A.: 3d mri brain tumor segmentation using autoencoder regularization. In: Brainlesion: Glioma, Multiple Sclerosis, Stroke and Traumatic Brain Injuries: 4th International Workshop, BrainLes 2018, Held in Conjunction with MICCAI 2018, Granada, Spain, September 16, 2018, Revised Selected Papers, Part II 4. pp. 311--320. Springer (2019)

\bibitem{oktay2018attention}
Oktay, O., Schlemper, J., Folgoc, L.L., Lee, M., Heinrich, M., Misawa, K., Mori, K., McDonagh, S., Hammerla, N.Y., Kainz, B., et~al.: Attention u-net: Learning where to look for the pancreas. arXiv preprint arXiv:1804.03999  (2018)

\bibitem{ragoza2023physics}
Ragoza, M., Batmanghelich, K.: Physics-informed neural networks for tissue elasticity reconstruction in magnetic resonance elastography. In: International Conference on Medical Image Computing and Computer-Assisted Intervention. pp. 333--343. Springer (2023)

\bibitem{ronneberger2015u}
Ronneberger, O., Fischer, P., Brox, T.: U-net: Convolutional networks for biomedical image segmentation. In: Medical Image Computing and Computer-Assisted Intervention--MICCAI 2015: 18th International Conference, Munich, Germany, October 5-9, 2015, Proceedings, Part III 18. pp. 234--241. Springer (2015)

\bibitem{selvaraju2017grad}
Selvaraju, R.R., Cogswell, M., Das, A., Vedantam, R., Parikh, D., Batra, D.: Grad-cam: Visual explanations from deep networks via gradient-based localization. In: Proceedings of the IEEE international conference on computer vision. pp. 618--626 (2017)

\bibitem{shanmugam2021better}
Shanmugam, D., Blalock, D., Balakrishnan, G., Guttag, J.: Better aggregation in test-time augmentation. In: Proceedings of the IEEE/CVF international conference on computer vision. pp. 1214--1223 (2021)

\bibitem{siren}
Sitzmann, V., Martel, J., Bergman, A., Lindell, D., Wetzstein, G.: Implicit neural representations with periodic activation functions. Advances in neural information processing systems  \textbf{33},  7462--7473 (2020)

\bibitem{song2020physics}
Song, T.A., Chowdhury, S.R., Yang, F., Jacobs, H.I., Sepulcre, J., Wedeen, V.J., Johnson, K.A., Dutta, J.: A physics-informed geometric learning model for pathological tau spread in alzheimer’s disease. In: Medical Image Computing and Computer Assisted Intervention--MICCAI 2020: 23rd International Conference, Lima, Peru, October 4--8, 2020, Proceedings, Part VII 23. pp. 418--427. Springer (2020)

\bibitem{zhang2019lookahead}
Zhang, M., Lucas, J., Ba, J., Hinton, G.E.: Lookahead optimizer: k steps forward, 1 step back. Advances in neural information processing systems  \textbf{32} (2019)

\end{thebibliography}

\end{document}